\documentclass[article,preprint,12pt,unsrt,compress]{elsarticle}
\usepackage{graphicx}
\usepackage{a4wide}
\usepackage{epstopdf}
\usepackage{booktabs}
\usepackage{multirow}
\usepackage{amssymb}
 \usepackage{amsthm}
 \usepackage{amsmath}
 \usepackage{hyperref}

\journal{
Physica A
}

\begin{document}

\begin{frontmatter}

\title{Has global warming modified the relationship between sunspot numbers and global temperatures?}

\author[ies]{Ladislav Kristoufek} \ead{ladislav.kristoufek@fsv.cuni.cz}

\address[ies]{Institute of Economic Studies, Faculty of Social Sciences, Charles University, Opletalova 26, 110 00, Prague 1, Czech Republic, EU} 

\begin{abstract}
We study time evolution of the relationship between sunspot numbers and global temperatures between 1880 and 2016 using wavelet coherence framework. The results suggest that the relationship is stable in time. Changes in the sunspot numbers precede changes in the temperatures by more than two years as suggested by the wavelet phase differences. This leading position of the sun activity is stable in time as well. However, the relationship has been disturbed by increasing $CO_2$ emissions since 1960s. Without controlling for the effect of possible global warming, or more precisely the positive connection between increasing $CO_2$ emissions and the global temperatures, the findings would have been quite different. Combination of the cointegration analysis and wavelet coherence framework has enabled uncovering a hidden relationship between the solar activity and global temperatures, and possibly explaining equivocal results in the topical literature.
\end{abstract}


\end{frontmatter}

\newpage

\section{Introduction}

Relationship between global climate and solar activity has been a topic of lively research activity for decades now \cite{Pittock1978,Pittock1979,Pittock1983,Loon2000}. Specifically the relationship between global temperatures and number of sunspots, as a proxy of solar activity\footnote{Note that sunspots number is not a perfect proxy of solar activity as documented in the literature \cite{Balmaceda2007,Scafetta2009,Hempelmann2012,Scafetta2014a}. However, we still use the proxy due to data availability and comparison with other studies reviewed later, which utilize number of sunspots as well.}, has started a new wave of discussion recently \cite{Scafetta2009,Rypdal2010,Scafetta2010,Gil-Alana2014,Scafetta2014,Gupta2015,Hassani2016}. The research question is usually not aimed simply to correlation between the two series but specifically whether the dynamics of solar activity precedes (leads or sometimes even causes) the temperature dynamics. A more detailed insight aims at potential non-linearity, structural changes in the relationship and cyclical behavior. In his early review, Pittock \cite{Pittock1978} summarizes that correlations are found on a scale of days but only weak ones at the level of characteristic sunspots cycles of 11 and 22 years while calling for a higher standard in statistical treatment of the issue. More than 30 years later, Pittock \cite{Pittock2009} is still rather pessimistic about the ability of solar activity to explain variation in global climate and global temperatures in particular mainly due to a problematic treatment of very long cycles in both sunspot numbers and global temperatures which, in addition, do not overlap. The cyclical properties of both time series are so important that most of the empirical analyses are based in the frequency domain. 

Controversies in the literature are nicely illustrated by two recent discussions. In the first one, Gil-Alana \textit{et al.} \cite{Gil-Alana2014} find that both series are fractionally integrated but of a different order and with a different dominant frequency by which they infer that this is a sing of no significant relationship between the two. Scafetta \cite{Scafetta2014} directly reacts to Ref. \cite{Gil-Alana2014} and opposes that these differences are primarily induced by non-stationarities and cyclical components of the series and argues that indeed there is a relationship between solar activity and surface temperatures, yet non-linear. This has lead to some further interest in testing non-linear causality between the series \cite{Gupta2015,Hassani2016}. And in the second one, Scafetta \& West \cite{Scafetta2003} identify corresponding L\'{e}vy components in both solar and temperature data, inferring a link between them. Rypdal \& Rypdal \cite{Rypdal2010} argue that when trends and moments scaling are taken into consideration, the link is gone. Scafetta \& West \cite{Scafetta2010} strike back arguing that the reasoning presented by Ref. \cite{Rypdal2010} is methodologically incorrect and they point to some further evidence based on spectral coherence \cite{Scafetta2010a}.

These disputes suggest several issues for this specific relationship. First, the results are strongly methodology-dependent. Second, cyclical components and trends play an essential role. Third, there are potential non-linearities in the link. And fourth, the relationship might be changing in time (the results differ for specific datasets). Here we propose to utilize a methodology that can overcome all the mentioned issues -- continuous wavelet coherence analysis. The methodology allows to study the relationship both in time and across frequencies in addition to its in-built robustness towards cyclical components and time trends. The following section gives a brief description of the wavelet coherence analysis. And the next two sections describe the analyzed dataset in detail and bring forward the results finding a structural break in the relationship between sunspot numbers and global temperature around year 1960 as well as discussing its connection to increasing emissions of carbon dioxide and temperatures in general in the same period of time. 

\section{Wavelet analysis and wavelet coherence}

Wavelet analysis can be used to decomposes a time series into several components with respect to their time and scale properties. Similarly to the Fourier analysis, wavelets can detect dominant cycles of the series at specific scales (frequencies). In addition to this scale perspective, wavelets broaden the decomposition and location into the time domain as well and they thus provide much richer information about dynamic properties of the analyzed series.

For time series $\{x_t\}$, wavelet $\psi_{u,s}(t)$ is a real or complex-valued function defined as 
\begin{equation}
\psi_{u,s}(t) =\frac{1}{\sqrt{s}} \psi {\left(\frac{t - u}{s}\right)},
\end{equation}
with a scale  parameter $s$ and a location parameter $u$. The original series $\{x_t\}$ can be fully recovered from its wavelet transform $W_x(u, s)$ defined as
\begin{equation}
W_x(u, s) = \int_{-\infty}^{+\infty}{x(t) \frac{1}{\sqrt{s}} \psi^{\ast} \left( \frac{t - u}{s}\right)\text{d}t},
\end{equation}
where $^\ast$ represents a complex conjugate operator preventing an information loss during the transformation \cite{Daub04}. The integral above measures the degree of similarity between the wavelet shape and  $\{x_t\}$. As wavelets can be either real or complex, wavelet power $|W_x(u,s)|^2$ is usually preferred in empirical applications as a parallel to variance. High levels of wavelet power identify dominant regions in the time-frequency space, i.e. dominant frequencies of the analyzed time series for a given time point.

The above described wavelet framework can be generalized into a bivariate (or in general multivariate) setting which allows to study connection between various time series and its evolution in time and across scales. Specifically for the bivariate setting, wavelet transform is generalized into cross-wavelet transform 
\begin{equation}
\label{xwt}
W_{xy}(u, s) = W_x(u, s) W_y^{\ast}(u, s),
\end{equation}
where $W_{x}(u, s)$ is wavelet transform of series $\{x_t\}$ and $W_y^{\ast}(u, s)$ marks a complex conjugate of wavelet transform of series $\{y_t\}$ \cite{Torr98}. Specifically, we utilize the Morlet wavelet $\psi^{M}(t)=\frac{1}{\pi^{1/4}}e^{i\omega_0t}e^{-t^2/2}$ with a central frequency $\omega_0 = 6$ which provides an optimal balance between time and frequency components \cite{Aguiar-Conraria2008,Rua2010}. As the Morlet wavelet, but also most of the cross-wavelet transforms, is complex, the cross-wavelet power is given by $|W_{xy}(u, s)|$ in the same line as the wavelet power $|W_x(u,s)|^2$. The cross-wavelet power is usually interpreted as a measure of local covariance between two series at a given frequency. However, its strength cannot be easily utilized for a detection of significant co-movement detection as it is not bounded. As this limitation is parallel to the limitation of standard covariance, the solution is again in the same logic. Analogously to correlation, the squared wavelet coherence is introduced as
\begin{equation}
R_{xy}^2(u, s) = \frac{|S\left(\frac{1}{s}W_{xy}(u, s)\right)|^2}{S\left(\frac{1}{s}|W_{x}(u, s)|^2 \right) S\left(\frac{1}{s}|W_{y}(u, s)|^2\right)},
\end{equation}
where $S$ is a smoothing operator \cite{Torr98b,Grin04}. The squared coherence $R_{xy}^2$ ranges between 0 and 1, and it corresponds to the usual squared correlation coefficient but here for specific time and frequency. Therefore, we get a value of squared coherence for each combination of a time point and scale. In practice, such correspondence is illustrated by a heat map where the squared coherence values are represented by different colors or shades for scaling purposes. In addition, the statistical significance of the coherence is tested using the Monte Carlo simulations, specifically, we test statistical significance against the null hypothesis of a red noise, i.e. an autoregressive process of order one. We return to this representation and more details in the Data and the Results sections when describing the results.

The imaginary part of the cross-wavelet transform $W_{xy}$ is not translated into the squared coherence due to the absolute value in the numerator of Eq. \ref{xwt}. The phase difference $\varphi_{xy}$, specified as     
\begin{equation}
\varphi_{xy}(u, s) = \text{tan}^{-1} \left(\frac{\mathfrak{I}\left[S\left(\frac{1}{s}W_{xy}(u, s)\right)\right]}{\mathfrak{R}\left[S\left(\frac{1}{s}W_{xy}(u, s)\right)\right]}\right)
\end{equation}
with $\mathfrak{R}$ and $\mathfrak{I}$ representing the real and imaginary part operators, respectively, helps retrieve this information. In practice, the phase is represented by an oriented arrow. For a specific time point and scale, an arrow pointing to the right suggests positive correlation between series (an in-phase relationship), and a left-oriented arrow suggests a negative correlation (an anti-phase relationship). A down-pointing arrow means that the first series leads the second series by $\pi/2$ and an upward-oriented arrow stands for the second series leading the first one by $\pi/2$. The orientations in between these four extremes are interpreted accordingly. We again return to the interpretation when analyzing the empirical data in the Results section.

\section{Data description}

We analyze the relationship between global temperatures and number of sunspots, specifically whether the changes in number of sunspots cause (or precede) changes in the temperatures. In addition, we are interested in possible time dependence of this dependence and its scale properties. Wavelet coherence framework described in the previous section provides an ideal environment for such analysis. For the temperatures, we use the dataset provided by the Goddard Institute of Space Studies (GISS) of NASA\footnote{Available at \url{http://data.giss.nasa.gov/gistemp}.}, specifically the global temperatures, and the temperatures for the northern and southern hemispheres separately. The time series are provided at monthly frequency between January 1880 and May 2016 (1637 observations) and they represent deviations from the 1951-1980 average with a scale of 0.01 $^\circ$C. Sunspot numbers series is obtained from the Solar Influence Data Analysis Centre (SIDC)\footnote{Available at \url{http://www.sidc.be/sunspot-data}.} and they cover monthly averages for the same period. 

\begin{figure}[!htbp]
\begin{center}
\begin{tabular}{c}
\includegraphics[width=160mm]{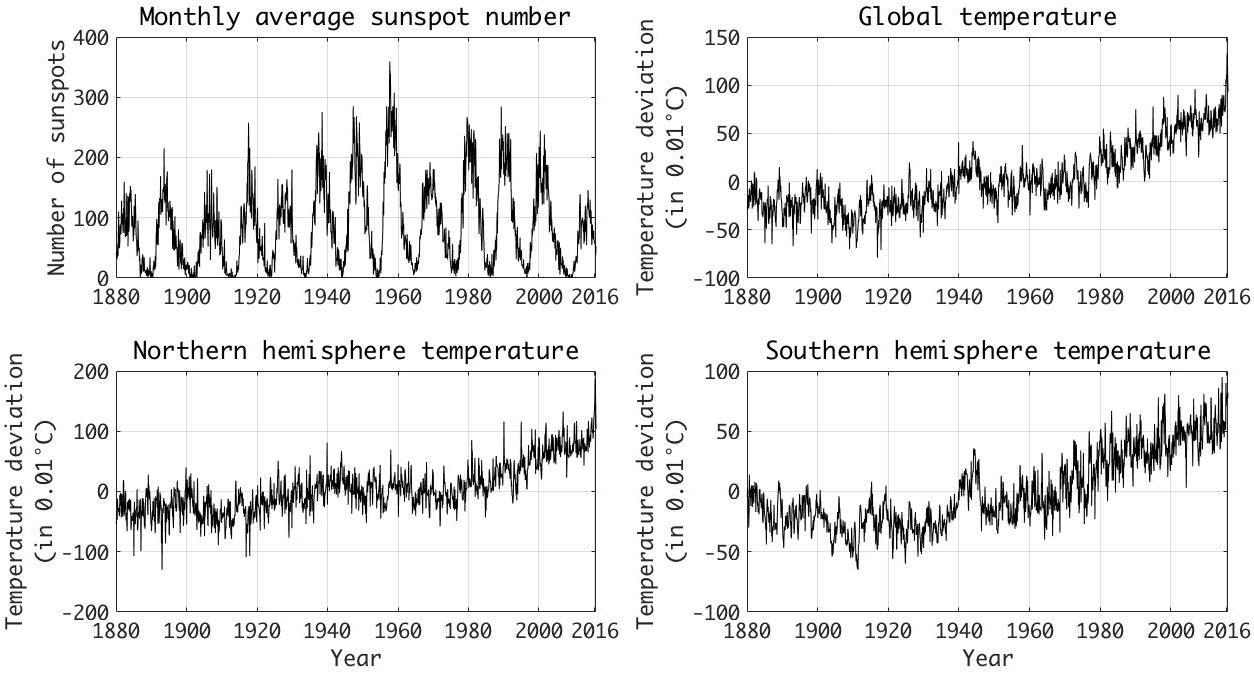}
\end{tabular}
\end{center}\vspace{-0.5cm}
\caption{\textbf{Sunspot numbers and global temperatures.} 
\footnotesize{The time series are shown between January 1880 and May 2016. Sunspot numbers are monthly averages of daily levels and temperatures are deviations from 1951-1980 average with a scale of 0.01 $^{\circ}$C. 
}
\label{Fig_Series}
}
\end{figure}

\begin{figure}[!htbp]
\begin{center}
\begin{tabular}{c}
\includegraphics[width=160mm]{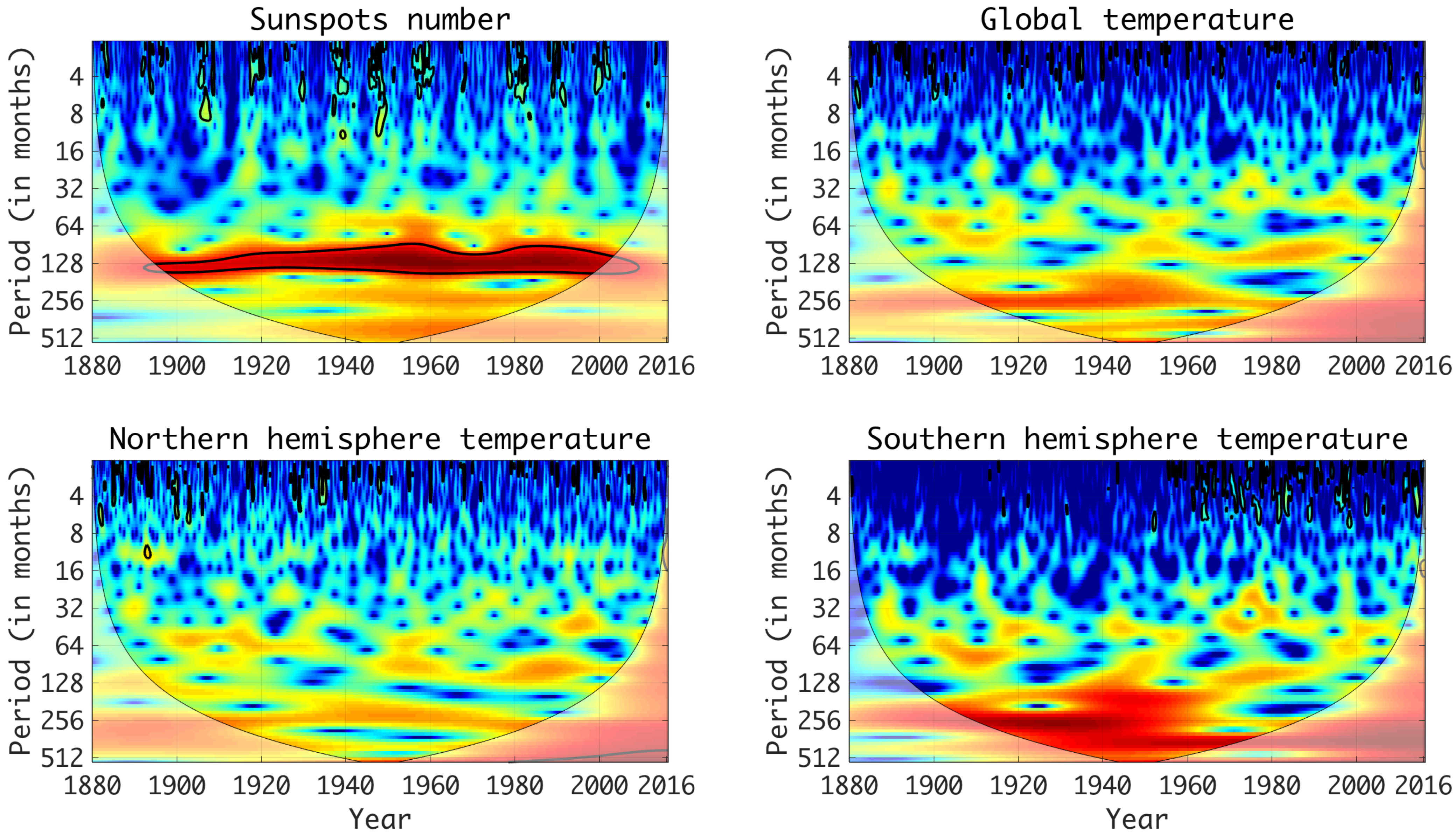}
\end{tabular}
\end{center}\vspace{-0.5cm}
\caption{\textbf{Wavelet powers.} 
\footnotesize{Wavelet powers are given for a given time point and a given scale (period). The level are represented by a color and the hotter the color the higher the power. Statistically significant regions are separated with a thick black curve. Cone of influence separates the reliable and unreliable regions of the time-scale plain -- full colors (north of the cone) represent the reliable area and pale colors (south of the cone) highlight the less reliable area. 
}
\label{Fig_Powers}
}
\end{figure}

The four analyzed series are illustrated in Fig. \ref{Fig_Series}. The sunspot numbers follow a strongly cyclical pattern with a varying amplitude but stable frequency. This is further supported in Fig. \ref{Fig_Powers} which shows wavelet powers for the analyzed series. The sunspot numbers show a dominant scale around 128 months, i.e. approximately 10 or 11 years, which is remarkably stable in time. This dominant cycle of sunspot numbers is in hand with results in the topical literature \cite{Pittock1978}. In the figure, the hotter the color, the higher the power. Thick black curve separates statistically significant and insignificant areas with respect to a null hypothesis of a red noise (i.e. an autocorrelated process with no dominant scale/frequency). The cone of influence represented by separating the time-scale domain into pale and full colors distinguishes a reliable (full colors) and an unreliable (pale colors) area of the domain due to wavelet stretching at high scales (low frequencies). The temperature series follow a similar pattern with respect to each other. They are all quite stable up till 1960 when an evident upward trend starts and keeps up until the end of the analyzed period. This upward trend is usually labelled as the global warming. We touch this later when discussing the results. The wavelet powers for the temperature series show no dominant scales, none of them is statistically significant even though we observe some large regions with high powers. This suggests two possibilities -- there either really is no dominant scale or the dominant scale is higher than 512 months (i.e. approximately 42 years), which cannot be detected by the wavelet analysis. Obviously only longer time series can answer this dilemma. Nevertheless, answering whether these upward trends are a sign of anthropomorphic global warming or a cyclical component with a very low frequency is not necessary for examining the relationship between the sunspot numbers and global temperatures \cite{Scafetta2013,Gervais2016}. Either way, these series have no common dominant scale in the analyzed range, i.e. below 512 months.

\section{Results}

\begin{figure}[!htbp]
\begin{center}
\begin{tabular}{c}
\includegraphics[width=160mm]{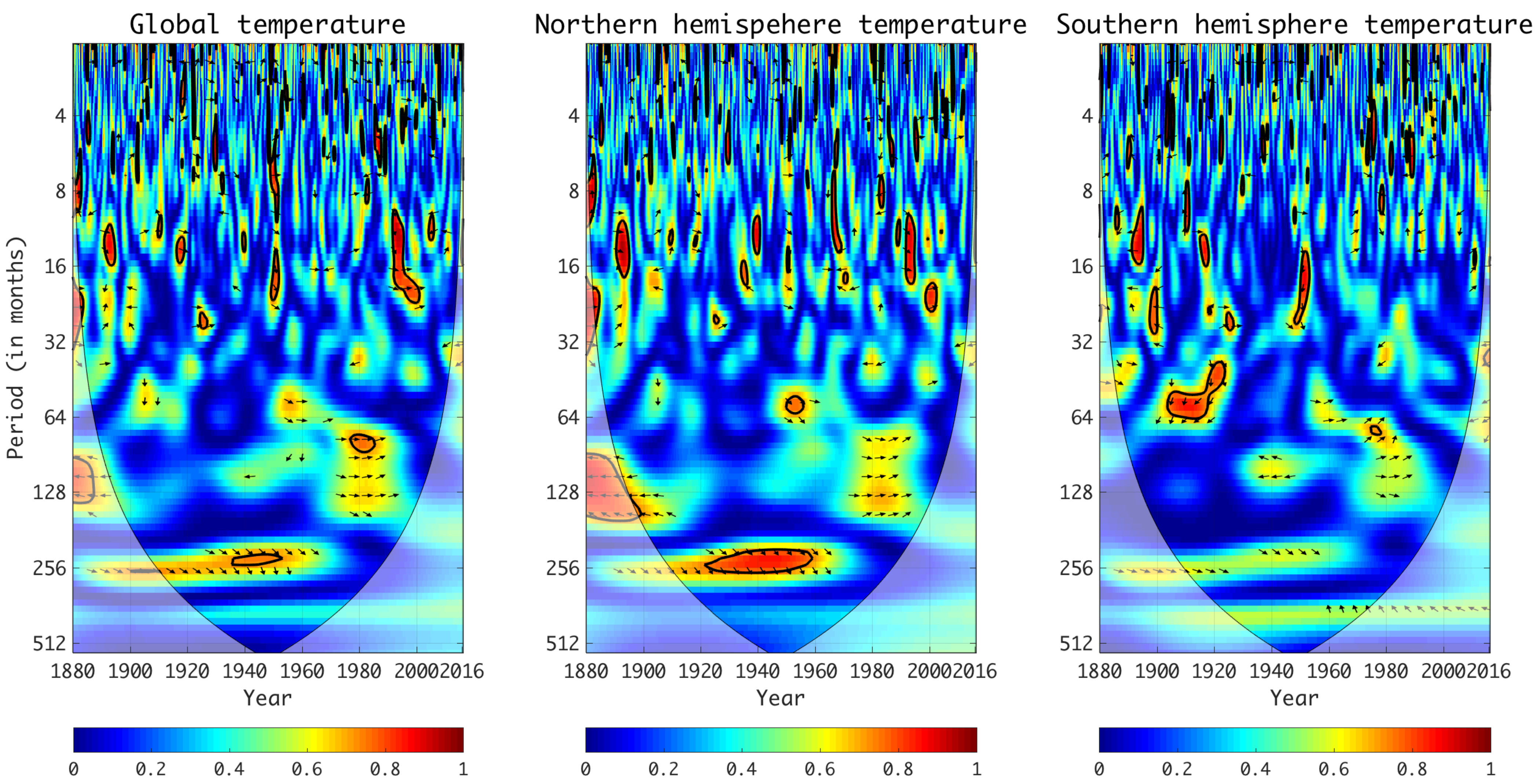}
\end{tabular}
\end{center}\vspace{-0.5cm}
\caption{\textbf{Squared wavelet coherence between sunspot numbers and global temperatures.} 
\footnotesize{The squared wavelet coherence is shown the pairs of the sunspot numbers and three temperature series between years 1880 and 2016. Values of the coherence are represented by a color map where the hot colors imply high coherence and cold colors imply low coherence (specific scale is shown at the bottom of each figure). Statistically significant regions, based on Monte Carlo simulations against the null hypothesis of the red noise, are marked by a thick black curve. The cone of influence separates the reliable (full colors) and less reliable (pale colors) regions. Phase difference arrows are shown for regions with the squared coherence above 0.5. 
}
\label{Fig_WTC}
}
\end{figure}

Looking at the wavelet coherence between sunspot numbers and temperatures in Fig. \ref{Fig_WTC}, we observe several interesting findings. First, the strength of dependence differs for the northern and southern hemisphere temperatures, it is stronger for the former and weaker for the latter. Second, the most prominent region with statistically significant wavelet coherence is identified at scales around 256 months (between 21 and 22 years). However, this relationship is significant only up to approximately 1960. Other regions are only short-lived and they can be attributed to noise. Note that this relationship is observed mainly for the northern hemisphere temperatures and it is practically gone for the southern hemisphere. Third, there is a straightforward interpretation of the phase relationship between the series. In Fig. \ref{Fig_WTC}, only arrows connected to wavelet coherence above 0.5 are shown. When looking at the significant region with the characteristic scale around 256 months, we observe that the arrows point southeast. This means that at this given scale, the sunspot number leads the temperature with a lead of approximately $\pi/4$, i.e. between 2 and 3 years, and these are positively correlated (in-phase). The increased solar activity, approximated by the number of sunspots, thus has a direct positive effect on the global temperatures, but with a pronounced lag. Such result is again more visible for the northern hemisphere temperatures.

\begin{figure}[!htbp]
\begin{center}
\begin{tabular}{c}
\includegraphics[width=100mm]{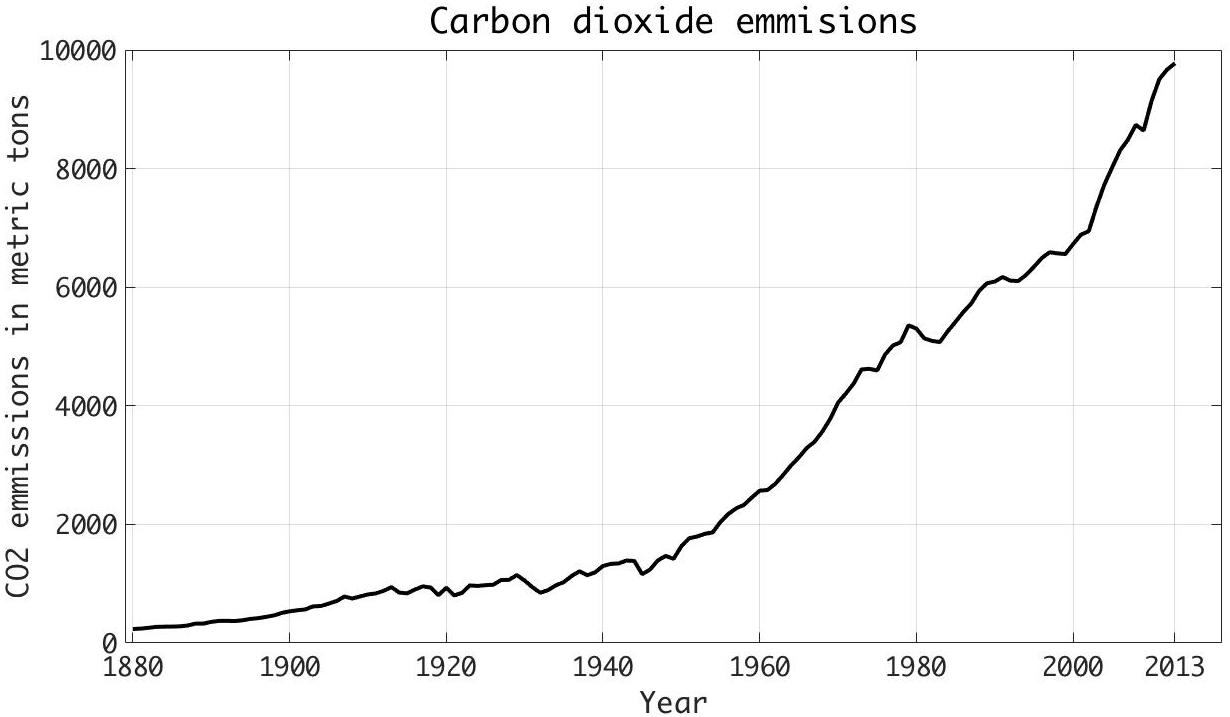} 
\end{tabular}
\end{center}\vspace{-0.5cm}
\caption{\textbf{Carbon dioxide emissions.} 
\footnotesize{Annual values in metric tons are shown for the period between 1880 and 2013.}
\label{Fig_CO2}
}
\end{figure}

The outcome of the wavelet coherence analysis is straightforward -- the solar activity measured by the sunspot numbers has a positive lagged effect on global temperatures. However, based on this bivariate analysis, such connection is observed only until approximately 1960. Looking back at Fig. \ref{Fig_Series}, this overlaps with the time period of quite stable global temperatures. However, when the temperatures start to increase, the dependence between sunspot numbers and temperatures seems to vanish. As the phenomenon of the global warming is usually connected with increasing concentrations of carbon dioxide ($CO_2$), we turn to this possible connection as well. Fig. \ref{Fig_CO2} shows the evolution of annual emissions of $CO_2$\footnote{Specifically global $CO_2$ emissions from fossil-fuel burning, cement manufacture, and gas flaring.} between years 1880 and 2013 based on the Carbon Dioxide Information Analysis Center database\footnote{Available at \url{http://cdiac.ornl.gov/trends/emis/tre_glob.html}.}. At the time of our analysis, the time series was available only up till 2013 and we proceed with a slightly reduced dataset hence. In the figure, we observe that $CO_2$ emissions are quite stable or slowly growing until approximately 1940. Since 1940, there has been a strong increasing trend which practically overlaps with the increasing global temperatures and weakening connection between the sunspot numbers and the temperatures. Controlling for the influence of $CO_2$ emissions could shed light on the true underlying relationship between sunspot numbers and global temperatures.

The time series of $CO_2$ emissions as well as the three temperature series are non-stationary (according to the Augmented Dickey-Fuller test \cite{Dick79} and the KPSS test \cite{Kwiatkowski1992}), which is well documented in the literature \cite{Galeotti2006,Richmond2006,Akbostanc2009,Fodha2010,Jaunky2011,Magazzino2016}, and they are even cointegrated\footnote{Testing statistics for (non-)stationarity and cointegration tests are available upon request.} (according to the Engle-Granger test \cite{Engle1987} and the Johansen test \cite{Johansen1991}). This implies that $CO_2$ and global temperatures tend to a common statistical equilibrium and return to common dynamics in a case of short-term deviations. Apart from the obvious interpretational importance, cointegration has crucial implications for estimation as well. For cointegrated series, we can apply standard least squares procedure which is super-consistent for this setting. We thus regress temperatures on $CO_2$ emissions and keep the residuals as deviations from the equilibrium relationship representing temperature dynamics after controlling for the effect of $CO_2$ emissions. This way, we can study the relationship between sunspot numbers and global temperatures as if there was no effect of $CO_2$ emissions.

\begin{figure}[!htbp]
\begin{center}
\begin{tabular}{c}
\includegraphics[width=160mm]{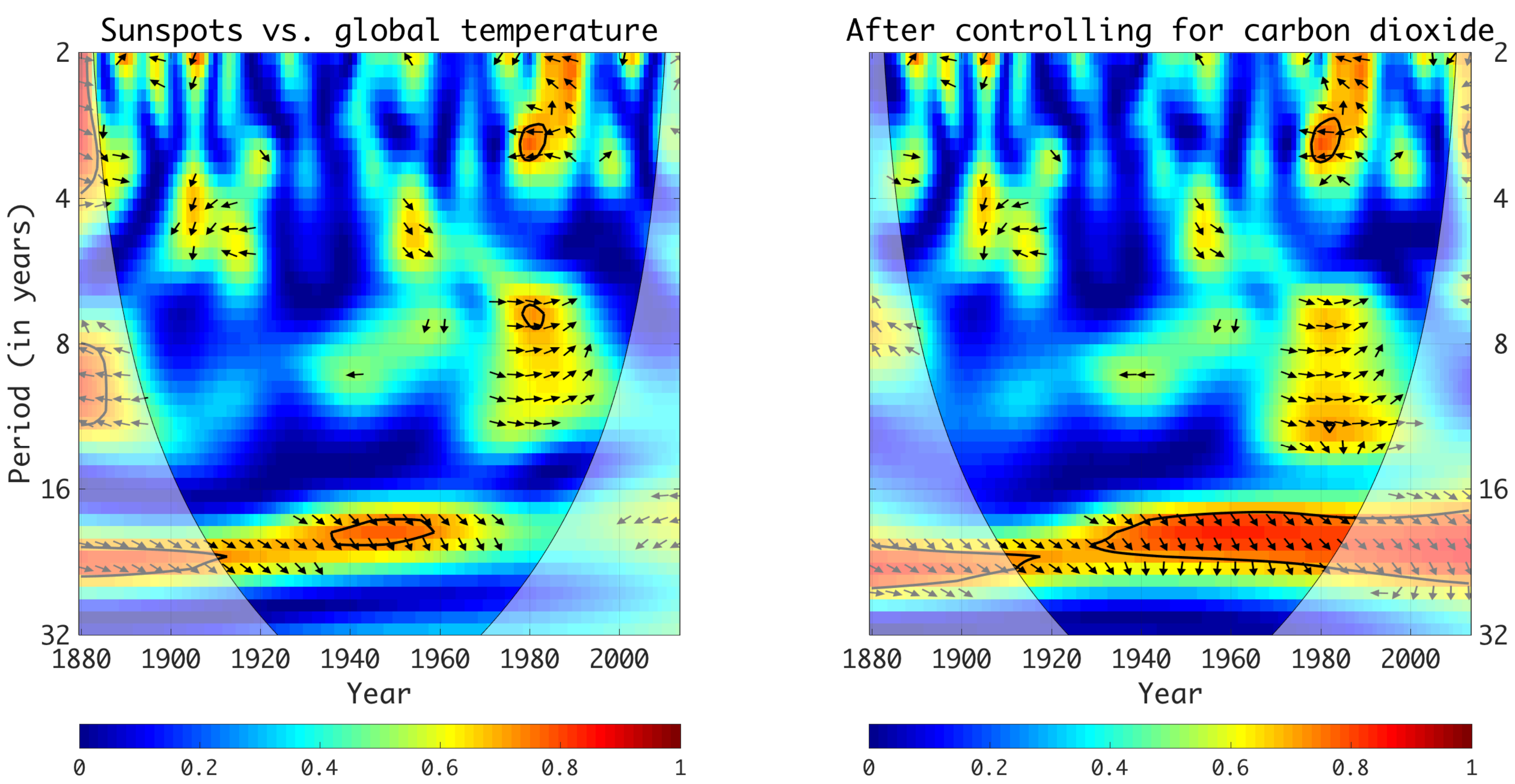}
\end{tabular}
\end{center}\vspace{-0.5cm}
\caption{\textbf{Squared wavelet coherence between sunspot numbers and global temperatures after controlling for the $CO_2$ effect.} 
\footnotesize{The left figure shows the results for original data with annual frequency. The right figure shows the relationship between the sunspot numbers and temperature corrected for the $CO_2$ effect using the cointegration framework. Other notations from Fig. \ref{Fig_WTC} hold here as well. The left figure copies the left part of Fig. \ref{Fig_WTC} but the right figure uncovers much stronger relationship between analyzed series. Not only are the significant regions larger but the connection between the two series is statistically significant even after 1960. This suggests that the $CO_2$ emissions increase after 1960 has adjusted the overall relationship between the global temperatures and sun activity.}
\label{Fig_CO2_Global}
}
\end{figure}

\begin{figure}[!htbp]
\begin{center}
\begin{tabular}{c}
\includegraphics[width=160mm]{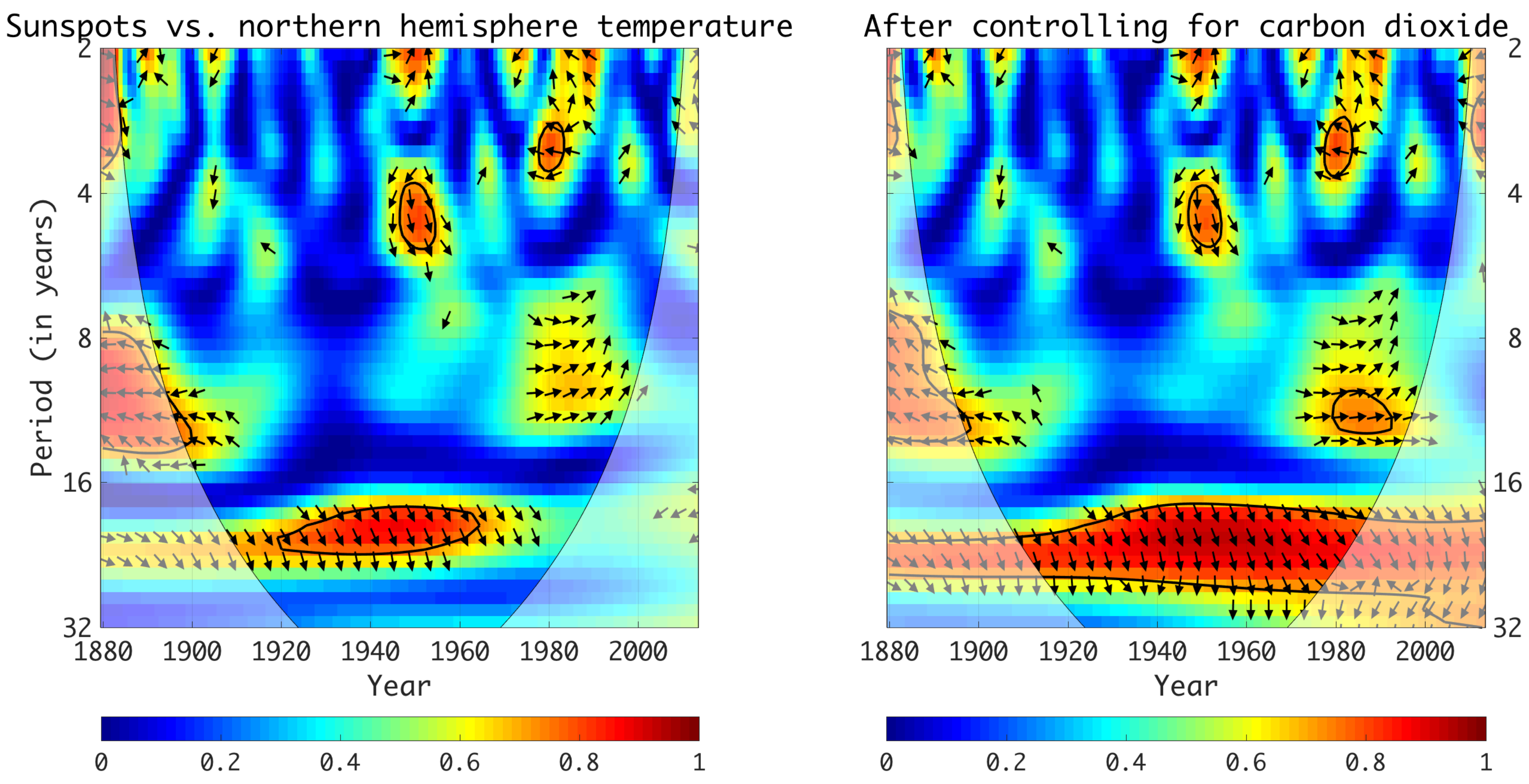}
\end{tabular}
\end{center}\vspace{-0.5cm}
\caption{\textbf{Squared wavelet coherence between sunspot numbers and northern hemisphere temperatures after controlling for the $CO_2$ effect.} 
\footnotesize{Notations from \ref{Fig_CO2_Global} hold here as well. Qualitatively, the results are very similar to the relationship between sunspot numbers and global temperatures. However, the connection is much stronger for the northern hemisphere.}
\label{Fig_CO2_Northern}
}
\end{figure}

\begin{figure}[!htbp]
\begin{center}
\begin{tabular}{c}
\includegraphics[width=160mm]{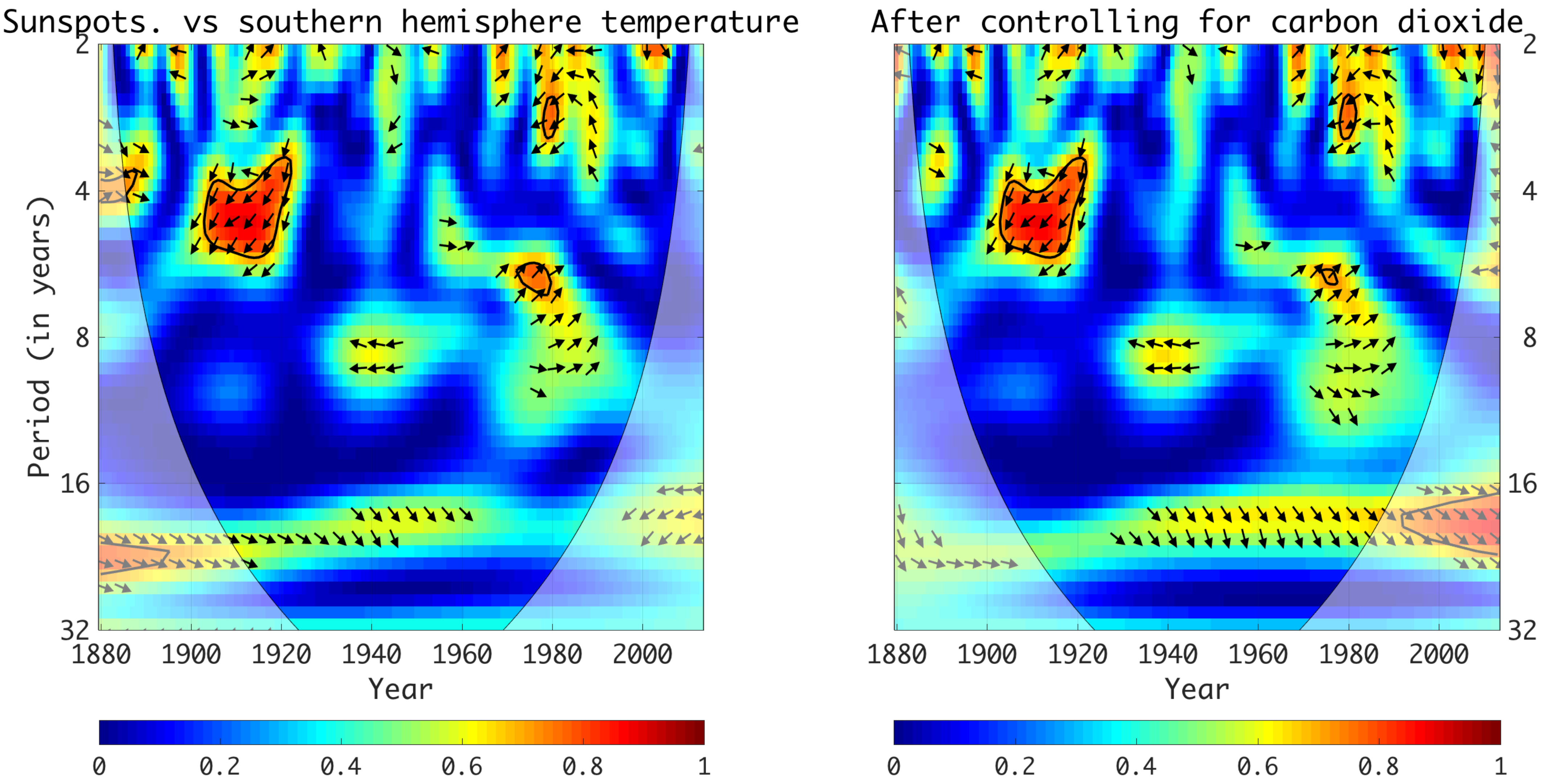}
\end{tabular}
\end{center}\vspace{-0.5cm}
\caption{\textbf{Squared wavelet coherence between sunspot numbers and southern hemisphere temperatures after controlling for the $CO_2$ effect.} 
\footnotesize{Notations from \ref{Fig_CO2_Global} hold here as well. Qualitatively, the results are very similar to the relationship between sunspot numbers and global temperatures. However, the connection is much weaker for the southern hemisphere and the connection is mostly insignificant.}
\label{Fig_CO2_Southern}
}
\end{figure}

The emissions time series has an annual frequency which implies that we need to lower the frequency for the sunspot numbers and temperatures as well. From the original series, we form annual averages of all four series of interest (sunspot numbers and three temperature series). Resulting wavelet coherences between the sunspot numbers and temperatures corrected for the effect of $CO_2$ emissions are shown in Figs. \ref{Fig_CO2_Global}-\ref{Fig_CO2_Southern}. As the frequency has been changed, we present the results both for the original temperatures and for the $CO_2$-corrected temperatures. The results are straightforward and strong. After controlling for the effect of the $CO_2$ emissions on temperatures, the dependence between sunspot numbers and temperatures increases considerably. Importantly, the connection is now statistically significant even for periods after 1960. The dominant scale overlaps with the one for original series, i.e. around 21-22 years. The correlation between series is positive and the changes in sunspot numbers precede the changes in temperatures, which is represented by phase arrows pointing southeast for the significant periods. In the same manner as for the original series, the relationship is much stronger for the northern hemisphere temperatures with statistically significant connection over the whole analyzed period (Fig. \ref{Fig_CO2_Northern}). Even though the coherence increased for the southern hemisphere as well, there is no statistically significant connection at scales above 16 years (Fig. \ref{Fig_CO2_Southern}). The presence of oriented arrows at scales above 16 years suggests that the squared wavelet coherence is above 0.5 there, i.e. strong but not statistically significant. The relationship for the global temperature (Fig. \ref{Fig_CO2_Global}) in a way represents an average between the southern and the northern hemisphere with statistically significant connection between the sunspot numbers and temperatures for almost the whole analyzed period. For the significant regions, the connection is positive and changes in the solar activity precede changes in the global temperatures. 

To provide a deeper insight into these results, we perform Granger causality tests in both time and frequency domains. In the time domain, the standard Granger methodology \cite{Granger1969} is utilized. In Table \ref{Time_Granger}, the results are summarized. These are presented for two types of relationships -- for the original series and for the ones with the $CO_2$ influence controlled for. There is no Granger causality between sunspot numbers and temperatures. This is not surprising as the standard time domain test does not take the scale (frequency) characteristics into consideration and thus practically averages the connection over all scales. Using methodology of Breitung \& Candelon \cite{Breitung2006,Croux2013}, we are able to test the Granger causality for specific scales or specific scale regions. Based on the results of the wavelet coherence analysis presented above, we split the scales into two regions -- below and above 16 years -- and test for Granger causality between sunspot numbers and temperatures for each region\footnote{We base the testing procedure on Croux \& Reusens \cite{Croux2013}, which is motivated by Breitung \& Candelon \cite{Breitung2006}. The procedure is based on setting the system of equations as a seemingly unrelated regression system, i.e. a system of equations with possibly correlated shocks. This setting fits perfectly for our dataset so that we can use three temperature variables as response variables and sunspot numbers as the impulse variable for all three equations. Used time lag is set to 11 years with respect to the frequency properties of the sunspot number series observable in Fig. \ref{Fig_Powers}.}. No Granger causality is found for scales below 16 years (with the testing statistics of 0.0064 and 0.0070 for original series and $CO_2$-corrected series, respectively, and the 90\% critical values of 0.0110 and 0.0109, respectively). For the scales above 16 years, the Granger causality is identified (with the testing statistics of 0.0127 and 0.0126 for original series and $CO_2$-corrected series, respectively, and the 90\% critical values of 0.0110 and 0.0109, respectively). The changes in sunspot numbers thus Granger-cause the changes in land temperatures for scales above 16 years.

\section{Conclusion}

In summary, the results suggest that the relationship between the solar activity, approximated by the sunspot numbers, and global temperatures is stable in time. Changes in the sunspot numbers precede changes in the temperatures by more than two years as suggested by the wavelet phase differences. This leading position of the solar activity is stable in time as well. However, this relationship has been disturbed by increasing $CO_2$ emissions since 1960s. Without controlling for the effect of global warming, or more precisely the positive effect of increasing $CO_2$ emissions on the global temperatures, the findings would have been different and, more importantly, biased. Combination of the cointegration analysis and wavelet coherence framework enabled uncovering a hidden relationship between the sunspot numbers and global temperatures.

\section*{Acknowledgements}

The research leading to these results was supported by the People Programme (Marie Curie Actions) of the European Union's Seventh Framework Programme FP7/2007-2013/ under REA grant agreement number 609642. The author further acknowledges financial support from the Czech Science Foundation (grants number 16-00027S).


\bibliographystyle{unsrt}

\begin{thebibliography}{10}

\bibitem{Pittock1978}
B.~Pittock.
\newblock A critical look at long-term sunñweather relationships.
\newblock {\em Reviews of Geophysics and Space Physics}, 16(3):400--420, 1978.

\bibitem{Pittock1979}
B.~Pittock.
\newblock Possible sunñweather correlation.
\newblock {\em Nature}, 280:254--255, 1979.

\bibitem{Pittock1983}
B.~Pittock.
\newblock Solar variability, weather and climate: an update.
\newblock {\em Quarterly Journal of the Royal Meteorological Society},
  109:23--55, 1983.

\bibitem{Loon2000}
H.~van Loon and K.~Labitzke.
\newblock The influence of the 11-year solar cycle on the stratosphere below 30
  km: A review.
\newblock {\em Space Science Reviews}, 94:259--278, 2000.

\bibitem{Balmaceda2007}
L.~Balmaceda, N.A. Krivova, and S.K. Solanki.
\newblock Reconstruction of solar irradiance using the {Group} sunspot number.
\newblock {\em Advances in Space Research}, 40:986--989, 2007.

\bibitem{Scafetta2009}
N.~Scafetta.
\newblock Empirical analysis of the solar contribution to global mean air
  surface temperature change.
\newblock {\em Journal of Atmospheric and Solar-Terrestrial Physics},
  71:1916--1923, 2009.

\bibitem{Hempelmann2012}
A.~Hempelmann and W.~Weber.
\newblock Correlation between the sunspot number, the total solar irradiance,
  and the terrestrial insolation.
\newblock {\em Solar Physics}, 277:417--430, 2012.

\bibitem{Scafetta2014a}
N.~Scafetta and R.C. Willson.
\newblock {ACRIM} total solar irradiance satellite composite validation versus
  {TSI} proxy models.
\newblock {\em Astrophysics and Space Science}, 350(2):421--442, 2014.

\bibitem{Rypdal2010}
M.~Rypdal and K.~Rypdal.
\newblock Testing hypotheses about sun-climate complexity linking.
\newblock {\em Physical Review Letters}, 104(12):128501, 2010.

\bibitem{Scafetta2010}
N.~Scafetta and B.J. West.
\newblock Comment on ``testing hypotheses about sun-climate complexity
  linking''.
\newblock {\em Physical Review Letters}, 105:219801, 2010.

\bibitem{Gil-Alana2014}
L.A. Gil-Alana, O.S. Yaya, and O.I. Shittu.
\newblock Global temperatures and sunspot numbers. are they related?
\newblock {\em Physica A}, 396:42--50, 2014.

\bibitem{Scafetta2014}
N.~Scafetta.
\newblock {Global temperatures and sunspot numbers. Are they related? Yes, but
  non linearly. A reply to {Gil-Alana} et al. (2014)}.
\newblock {\em Physica A}, 413:329--342, 2014.

\bibitem{Gupta2015}
R.~Gupta, L.A. Gil-Alana, and O.S. Yaya.
\newblock Do sunspot numbers cause global temperatures? evidence from a
  frequency domain causality test.
\newblock {\em Applied Economics}, 47(8):798--808, 2015.

\bibitem{Hassani2016}
H.~Hassani, X.~Huang, R.~Gupta, and M.~Ghodsi.
\newblock Does sunspot numbers cause global temperatures? a reconsideration
  using non-parametric causality tests.
\newblock {\em Physica A}, 460:54--65, 2016.

\bibitem{Pittock2009}
B.~Pittock.
\newblock Can solar variations explain variations in the {E}arth's climate?
\newblock {\em Climate Change}, 96:483--487, 2009.

\bibitem{Scafetta2003}
N.~Scafetta and B.J. West.
\newblock Solar flare intermittency and the {E}arth's temperature anomalies.
\newblock {\em Physical Review Letters}, 90(24):248701, 2003.

\bibitem{Scafetta2010a}
N.~Scafetta.
\newblock Empirical evidence for a celestial origin of the climate oscillations
  and its implications.
\newblock {\em Journal of Atmospheric and Solar-Terrestrial Physics},
  72:951--970, 2010.

\bibitem{Daub04}
Ingrid Daubechies.
\newblock {\em Ten Lectures on Wavelets}.
\newblock SIAM, 2004.

\bibitem{Torr98}
Christopher Torrence and Gilbert~P. Compo.
\newblock A practical guide to wavelet analysis.
\newblock {\em Bulletin of the American Meteorological Society}, 79:61--78,
  1998.

\bibitem{Aguiar-Conraria2008}
L.~Aguiar-Conraria, L.~Azevedo, and M.~Soares.
\newblock Using wavelets to decompose the time-frequency effects of monetary
  policy.
\newblock {\em Physica A}, 387:2863--2878, 2008.

\bibitem{Rua2010}
A.~Rua.
\newblock Measuring comovement in the time-frequency space.
\newblock {\em Journal of Macroeconomics}, 32:685--691, 2010.

\bibitem{Torr98b}
C.~Torrence and P.J. Webster.
\newblock The annual cycle of persistence in the {El Nino-Southern
  Oscillation}.
\newblock {\em Quarterly Journal of the Royal Meteorological Society},
  124(550):1985--2004, 1998.

\bibitem{Grin04}
Aslak Grinsted, {J C} Moore, and S~Jevrejeva.
\newblock Application of the cross wavelet transform and wavelet coherence to
  geophysical time series.
\newblock {\em Nonlinear Processes in Geophysics}, 11:561--566, 2004.

\bibitem{Scafetta2013}
N.~Scafetta.
\newblock Discussion on climate oscillations: {CMIP5} general circulation
  models versus a semiempirical harmonic model based on astronomical cycles.
\newblock {\em Earth-Science Reviews}, 126:321--357, 2013.

\bibitem{Gervais2016}
F.~Gervais.
\newblock Anthropogenic $co_2$ warming challenged by 60-year cycle.
\newblock {\em Earth-Science Reviews}, 155:129--135, 2016.

\bibitem{Dick79}
David~A Dickey and Wayne~A Fuller.
\newblock Distribution of the estimators for autoregressive time series with a
  unit root.
\newblock {\em Journal of the American statistical association}, 74:427--431,
  1979.
\newblock Taylor \& Francis.

\bibitem{Kwiatkowski1992}
D.~Kwiatkowski, P.~Phillips, P.~Schmidt, and Y.~Shin.
\newblock Testing the null of stationarity against alternative of a unit root:
  How sure are we that the economic time series have a unit root?
\newblock {\em Journal of Econometrics}, 54:159--178, 1992.

\bibitem{Engle1987}
R.~F. Engle and C.~W.~J. Granger.
\newblock Co-integration and error correction: Representation, estimation, and
  testing.
\newblock {\em Econometrica}, 55:251--276, 1987.

\bibitem{Johansen1991}
S.~Johansen.
\newblock Estimation and hypothesis testing of cointegration vectors in
  gaussian vector autoregressive models.
\newblock {\em Econometrica}, 59(6):1551--1580, 1991.

\bibitem{Galeotti2006}
M.~Galeotti, A.~Lanza, and F.~Pauli.
\newblock Reassessing the environmental kuznets curve for $co_2$ emissions. a
  robustness exercise.
\newblock {\em Econological Economy}, 57:452--456, 2006.

\bibitem{Richmond2006}
A.~Richmond and R.~Kaufmann.
\newblock Is there a turning point in the future relationship between income
  and energy use and or carbon emissions?
\newblock {\em Ecological Economics}, 56:176--186, 2006.

\bibitem{Akbostanc2009}
E.~Akbostanc, S.~Turut, and G.~Tunc.
\newblock The relationship between income and environment in {Turkey}: is there
  an environmental kuznet's curve?
\newblock {\em Economic Policy}, 37:861--867, 2009.

\bibitem{Fodha2010}
M.~Fodha and O.~Zaghdoud.
\newblock Economic growth and pollutant emissions in tunisia. an empirical
  analysis of the environmental kuznet's curve.
\newblock {\em Energy Policy}, 38:1150--1156, 2010.

\bibitem{Jaunky2011}
V.~Jaunky.
\newblock The $co_2$ emissions income nexus: Evidence from rich countries.
\newblock {\em Energy Policy}, 39:1228--1234, 2011.

\bibitem{Magazzino2016}
C.~Magazzino.
\newblock The relationship between $co_2$ emissions and energy consumption and
  economic growth in {Italy}.
\newblock {\em International Journal of Sustainable Energy}, 35(9):844--857,
  2016.

\bibitem{Granger1969}
C.W.J. Granger.
\newblock Investigating causal relations by econometrics models and
  cross-spectral methods.
\newblock {\em Econometrica}, 37(3):424--438, 1969.

\bibitem{Breitung2006}
J.~Breitung and B.~Candelon.
\newblock Testing for short- and long-run causality: {A frequency-domain
  approach}.
\newblock {\em Journal of Econometrics}, 132:363--378, 2006.

\bibitem{Croux2013}
C.~Croux and P.~Reusens.
\newblock Do stock prices contain predictive power for the future economic
  activity? a granger causality analysis in the frequency domain.
\newblock {\em Journal of Macroeconomics}, 35:93--103, 2013.

\end{thebibliography}

\newpage

\begin{table}[!htbp]
\caption{\textbf{Time-domain Granger causality tests.} Causality from sunspot numbers towards temperatures is tested. The lag number is based on the Schwarz-Bayesian information criterion (SBIC). $p$-values over 0.1 suggest there is no Granger causality between the series.\label{Time_Granger}}
\centering
\begin{tabular}{|c||c|c|c||c|c|c|}
\hline \hline
&testing statistic&\multirow{2}{*}{$p$-value}&\multirow{2}{*}{lags}&testing statistic&\multirow{2}{*}{$p$-value}&\multirow{2}{*}{lags}\\
&(original)&&&($CO_2$ controlled)&&\\
\hline
sunspots $\rightarrow$ global $T$&0.1419&0.9347&3&0.4734&0.6240&2\\
sunspots $\rightarrow$ northern $T$&0.0844&0.9191&2&0.1383&0.8710&2\\
sunspots $\rightarrow$ southern $T$&0.1524&0.9280&3&0.7316&0.4832&2\\
\hline \hline
\end{tabular}
\end{table}

\end{document}